\documentclass[11pt]{article}

\usepackage{amsmath,amssymb,latexsym,color}
\usepackage{pstricks}
\usepackage{multicol}
\usepackage{graphicx}
\usepackage{amsfonts}

\newsavebox{\astrutbox}
\sbox{\astrutbox}{\rule[-5pt]{0pt}{20pt}}

\providecommand\bnabla{\boldsymbol{\nabla}}

\newcommand\p{\ensuremath{\partial}}

\newcommand\shalf{\ensuremath{{\scriptstyle\frac{1}{2}}}}

\newcommand\etal{\mbox{\textit{et al. }}}

\newcommand{\rem}[1]{}
\newcommand{\remfigure}[1]{}
\newtheorem{theorem}{Theorem}

\DeclareMathAlphabet{\mathbi}{OML}{cmm}{b}{it} 
\newcommand{\bx}{\mathbi{x}}
\newcommand{\bX}{\mathbi{X}}
\newcommand{\by}{\mathbi{y}}
\newcommand{\bel}{\begin{equation}\label}
\newcommand{\ee}{\end{equation}}
\newcommand{\ben}{\begin{enumerate}}
\newcommand{\een}{\end{enumerate}}
\newcommand{\bB}{\mbox{\boldmath$B$}}
\newcommand{\bhB}{\mbox{\boldmath$\hat{B}$}}
\newcommand{\bdb}{\mbox{\boldmath$b$}}
\newcommand{\bD}{\mbox{\boldmath$\mathcal{D}$}}

\newcommand{\ba}{\mathbi{a}}

\newcommand{\br}{\mathbi{r}}
\newcommand{\bs}{\mathbi{s}}
\newcommand{\bq}{\mathbi{q}}
\newcommand{\be}{\mathbi{e}}
\newcommand{\bqr}{\mbox{\boldmath$\omega_{\rho}$}}
\newcommand{\bqt}{\mbox{\boldmath$\tilde{\omega}$}}

\newcommand{\bhn}{\mathbi{\hat{n}}}
\newcommand{\bfq}{\mathfrak{q}}
\newcommand{\bhfq}{\hat{\mathfrak{q}}}
\newcommand{\bhfp}{\hat{\mathfrak{p}}}

\newcommand{\bfw}{\mathfrak{w}}

\newcommand{\bfr}{\mathfrak{r}}
\newcommand{\cast}{\circledast}

\newcommand{\bmu}{\mbox{\boldmath$\mu$}}

\newcommand{\bdl}{\mbox{\boldmath$\delta\ell$}}
\newcommand{\bu}{\mathbi{u}}
\newcommand{\bv}{\mathbi{v}}

\newcommand{\bQ}{\mathbi{Q}}
\newcommand{\bw}{\mathbi{w}}
\newcommand{\bhw}{\mathbi{\hat{w}}}
\newcommand{\bW}{\mathbi{W}}

\newcommand{\bPi}{\mathfrak{P}}
\newcommand{\bom}{\mbox{\boldmath$\omega$}}
\newcommand{\bhom}{\mbox{\boldmath$\hat{\omega}$}}
\newcommand{\bcapom}{\mbox{\boldmath$\Omega$}}

\newcommand{\bchi}{\mbox{\boldmath$\chi$}}
\newcommand{\bhchi}{\boldsymbol{\hat{\chi}}}

\newcommand{\beq}{\begin{eqnarray}} 
\newcommand{\eeq}{\end{eqnarray}} 
\newcommand{\bc}{\begin{center}} 
\newcommand{\ec}{\end{center}} 
\newcommand{\lin}{L^{\infty}(\mathbb{D})}

\makeatletter
\@addtoreset{equation}{section}

\makeatother

\begin{document}
\bc
\textbf{\large Ortho-normal quaternion frames, Lagrangian evolution equations and the 
three-dimensional Euler equations\footnote{The material in this review is based on the 
contents of an invited lecture given at the meeting \textit{Mathematical Hydrodynamics} 
held at the Steklov Institute Moscow, in June 2006.}}
\ec

\bc
\textbf{\large J. D. Gibbon}
\par\medskip
Department of Mathematics, Imperial College London\footnote{email address: j.d.gibbon@ic.ac.uk}, 
London SW7 2AZ, UK
\ec

\date{3rd November, 2006}

\begin{abstract}
More than 160 years after their invention by Hamilton, quaternions are now widely used 
in the aerospace and computer animation industries to track the orientation and paths 
of moving objects undergoing three-axis rotations. It is shown here that they provide 
a natural way of selecting an appropriate ortho-normal frame -- designated the 
quaternion-frame -- for a particle in a Lagrangian flow, and of obtaining the 
equations for its dynamics. How these ideas can be applied to the three-dimensional 
Euler fluid equations is then considered. This work has some bearing on the issue of 
whether the Euler equations develop a singularity in a finite time. Some of the literature 
on this topic is reviewed, which includes both the Beale-Kato-Majda theorem and associated 
work on the direction of vorticity by Constantin, Fefferman \& Majda and Deng, Hou 
and Yu. It is then shown how the quaternion formulation provides an alternative formulation 
in terms of the Hessian of the pressure. 
\end{abstract}

\vspace{5mm}

\bc
\textit{This paper is dedicated to the memory of Victor Yudovich (1934-2006) with whom 
the author discussed some of these ideas in their early stages.}
\ec

{\small\tableofcontents}

\
\section{\large General introduction}\label{Gintro}

\subsection{Historical remarks}\label{Hintro}

Everyone loves a good story\,: William Rowan Hamilton's feverish excitement at the discovery 
of his famous formula for quaternions on 16th October 1843 as a composition rule for 
orienting his telescope; his inscription of this formula on Broome (Brougham) Bridge in 
Dublin; and then his long and eventually unfruitful championing of the role of quaternions 
in mechanics, are all elements of a story that has lost none of its appeal \cite{OR1998,Ha2006}. 
Hamilton's name is still revered today for the audacity and depth of his ideas in modern 
mechanics and what we now call symplectic geometry \cite{Arnold1,Mars81,Mars92}. Indeed, 
evidence of his thinking is everywhere in both classical and quantum mathematical physics 
and applied mathematics, 
yet in his own century his work on quaternions evoked criticism and even derision from many 
influential fellow scientists\footnote{Kelvin was one such example: see \cite{OR1998}.}. 
Ultimately quaternions lost out to the tensor notation of Gibbs, which is the basis of the 
3-vector notation universally used today. 
\par\smallskip
In essence, Hamilton's multiplication rule for quaternions represents compositions of 
rotations \cite{Cay1845,Ham1853,Ham1866,Ta1890,Whitt1944,Klein04}. This property has been 
ably exploited in modern inertial guidance systems in the aerospace industry where computing 
the orientation and the paths of rapidly moving rotating satellites and aircraft is essential. 
Kuipers' book explains the details of how calculations with quaternions in this field are 
performed in practice \cite{Ku1999}. Just as importantly, the computer graphics community also 
uses them to determine the orientation of tumbling objects in animations. In his valuable and 
eminently readable book, Andrew Hanson \cite{Ha2006} says in his introduction\,:
\begin{quote}{\small
Although the advantages of the quaternion forms for the basic equations of attitude control -- 
clearly presented in Cayley \cite{Cay1845}, Hamilton \cite{Ham1853,Ham1866} and especially 
Tait \cite{Ta1890} -- had been noticed by the aeronautics and astronautics community, the 
technology did not penetrate the computer animation community until the land-mark Siggraph 
1985 paper of Shoemake \cite{Shoe1985}. The importance of Shoemake's paper is that it took 
the concept of the orientation frame for moving $3D$ objects and cameras, which require 
precise orientation specification, exposed the deficiencies of the then-standard Euler-angle 
methods\footnote{A well-known deficiency of Euler-angle methods lies in the problems they 
suffer at the poles of the sphere where the azimuthal angle is not defined.}, and introduced 
quaternions to animators as a solution.}
\end{quote}
Hamilton's 19th century critics were, of course, correct in their assertion that quaternions 
need 3-vector algebra to manipulate them, yet the use the aero/astronautics and animation 
communities have made of them are one more illustration of the universally acknowledged truth 
that while new mathematical tools may not be of immediate use, and may appear to be too 
abstract or overly elaborate, they may nevertheless turn out to have powerful applications 
undreamed of at the time of their invention.

\subsection{Application to fluid dynamics}\label{applic}

The close association of quaternions with rigid body rotations \cite{Ta1890,Whitt1944,Klein04} 
points to their use in the incompressible Euler equations for an inviscid fluid as a natural 
language for describing the alignment of vorticity with the eigenvectors of the strain rate 
that are responsible for its nonlinear evolution. For a three-dimensional fluid velocity field 
$\bu(\bx,\,t)$ with pressure $p(\bx,\,t)$, the incompressible Euler equations are 
\cite{Eul1755,MB01,GKB,Saffbk,MKO}
\bel{eul1}
\frac{D\bu}{Dt} = -\nabla p\,,
\ee
where the material derivative is defined by
\bel{mat1}
\frac{D~}{Dt} = \frac{\partial~}{\partial t} + \bu\cdot \nabla\,.
\ee
The motion is constrained by the incompressibility condition $\hbox{div}\,\bu = 0$. 
The crucial dynamics lies in the evolution of the velocity gradient matrix 
$\nabla\bu = \{u_{i,j}\}$ which comes from the differentiation of (\ref{eul1})
\bel{vgm1}
\frac{D u_{i,j}}{Dt} = - u_{i,k}u_{k,j} - P_{ij}\,,
\ee
where $P_{ij}$ is the Hessian matrix of the pressure 
\bel{Hessdef}
P_{ij} = \frac{\partial^{2}p}{\partial x_{i}\partial x_{j}}\,.
\ee
The incompressibility condition $\hbox{div}\,\bu = 0$ insists that $Tr\,u_{i,j} = 0$ which, 
when applied to (\ref{vgm1}), gives 
\bel{Trace}
Tr\,P = \Delta p = - u_{i,k}u_{k,i} = \shalf\omega^{2} - Tr\,(S^{2})\,.
\ee
In (\ref{Trace}) above, $S$ is the strain matrix whose elements are defined by
\bel{straindef}
S_{ij} = \shalf\left(u_{i,j} + u_{j,i}\right)\,.
\ee 
This is a symmetric matrix the alignment of whose eigenvectors $\be_{i}$ is fundamental 
to the dynamics of the Euler equations. For instance, vortex tubes and sheets (Burgers' 
vortices and shear layers) always have one eigenvector aligned with the vorticity vector 
$\bom$ \cite{MKO}.
\par\medskip
This review cannot hope to deal with every aspect of the three-dimensional Euler equations, 
particularly the vast literature on weak and distributional solutions; the reader is urged 
to read the book by Majda \& Bertozzi \cite{MB01} to study these aspects of the problem. 
Here we concentrate on one particular aspect, which is the role played by quaternions  
in providing a natural language for extracting geometric information from the evolution of 
$u_{i,j}$. Because they are particularly effective in computing the orientation of rotating 
objects moving in three-dimensional paths they might be useful in understanding how general 
Lagrangian flows behave, particularly in finding the evolution of the ortho-normal frame 
of particles moving in such a flow. These particles could be of the passive tracer type 
transported by a background flow or they could be Lagrangian fluid parcels. 
Recent experiments in turbulent flows can now detect the trajectories of tracer particles 
at high Reynolds numbers 
\cite{La01,M01,Vo02,MCB04,MLP04,M05,LTK05,Bif05,RMCB05,Eck06}\,: see Figure 1 in \cite{La01}.  
For any system involving a path represented as a three-dimensional space-curve, the usual practice 
is to consider the Frenet-frame of a trajectory constituted by the unit tangent vector, the 
normal and the bi-normal \cite{Ha2006,Eck06}. In navigational language, this represents the 
corkscrew-like pitch, yaw and roll of the motion.  While the Frenet-frame describes the path, 
it ignores the dynamics that generates the motion. Attempts have been made in this direction 
using the eigenvectors $\be_{i}$ of $S$ but ran into difficulties because the equations of 
motion for $\be_{i}$ are unknown \cite{Tabor}.  In \S\ref{lagev} another ortho-normal frame 
is introduced that is associated with the motion of each Lagrangian particle. 
It is designated the \textit{quaternion-frame}\,: this frame may be envisioned as moving with 
the Lagrangian particles, but its evolution derives from the Eulerian equations of motion.  The 
advantage of this approach lies in the fact that the Lagrangian dynamics of the quaternion-frame 
can be connected to the fluid motion through the pressure Hessian $P$ defined in (\ref{Hessdef}). 
\par\smallskip
Let us now consider a general picture of a Lagrangian flow system of equations. 
Suppose $\bw$ is a contravariant vector quantity attached to a particle following a flow 
along characteristic paths $d\bx/dt=\bu(\bx,\,t)$ of a velocity field $\bu$. Let us consider 
the abstract Lagrangian flow equation
\bel{w-dyn1}
\frac{D\bw}{Dt} = \ba(\bx,\,t)\,,
\hspace{3cm}
\frac{D~}{Dt} = \frac{\p~}{\p t} + \bu\cdot\nabla\,,
\ee
where the material derivative has its standard definition, and that, in turn, $\ba$ satisfies 
the Lagrangian equation
\bel{a-dyn2}
\frac{D^2\bw}{Dt^2} = \frac{D\ba}{Dt} = \bdb(\bx,\,t)\,.
\ee
So far, these are just kinematic rates of change following the characteristics of the velocity 
generating the path $\bx(t)$ determined from $d\bx/dt=\bu(\bx,\,t)$. Examples of systems that 
(\ref{w-dyn1}) might represent are\,:
\ben\itemsep -1mm

\item If $\bw$ represents the vorticity $\bom = \hbox{curl}\,\bu$ of the incompressible 
Euler fluid equations then $\ba = \bom\cdot\nabla\bu$ and $\hbox{div}\,\bu = 0$. With rotation 
$\bw$ would be $\bw\equiv\bqt = \rho_{0}^{-1}(\bom + 2\bcapom)$.

\item For the barotropic compressible Euler fluid equations (where the pressure $p = p(\rho)$ 
is density dependent only) then $\bw \equiv \bqr = \rho^{-1}\bom$, in which case 
$\ba = \bqr\cdot\nabla\bu$ and $\hbox{div}\,\bu = 0$. 

\item $\bw$ could also represent a small vectorial line element $\bdl$ that is mixed and 
stretched by a background flow $\bu$, in which case $\ba = \bdl\cdot\nabla\bu$. For example, 
following Moffatt's analogy with between the magnetic field $\bB$ in ideal incompressible  
MHD and vorticity \cite{HKM1}, if $\bw$ is chosen such that $\bw \equiv \bB$, then 
$\ba = \bB\cdot\nabla\bu$ with $\hbox{div}\,\bB = 0$. In a more generalized form it could 
also represent the Elsasser variables $\bw^{\pm} = \bu \pm \bB$, in which case $\ba^{\pm} = 
\bw^{\pm}\cdot\nabla\bu$ with two material derivatives.

\item The semi-geostrophic (SG) model used in atmospheric physics can also be cast in the form 
of (\ref{w-dyn1})\,; for instance one could choose $\bw =\bx$, $\ba = \bu$ and $\bdb$ 
is computed from the SG-model through the semi-geostrophic and a-geostrophic contributions 
\cite{NorRoul02,Majdaatmos03,Cullen06}.

\item For a passive tracer particle with velocity $\bw$ in a fluid transported by a background 
velocity field $\bu$, the particle's acceleration would be $\ba$ (see \cite{FaGaVe2001,GKB}). 
\een
In cases (1--3) above if $\bw$ satisfies the standard Eulerian form
\bel{wev1}
\frac{D\bw}{Dt} = \bw\cdot\nabla\bu \,,
\ee
then to find $\bdb$ it follows from Ertel's Theorem that \cite{Er42}
\bel{ertel2}
\frac{D(\bw\cdot\nabla\bmu)}{Dt} = \bw\cdot\nabla\left(\frac{D\bmu}{Dt}\right)\,,
\ee
which means that the operators $D/Dt$ and $\bw\cdot\nabla$ commute for any differentiable function 
$\bmu(\bx,\,t)$. Choosing $\bmu = \bu$ as in \cite{Ohk93}, and identifying the flow 
acceleration as $\bQ(\bx,\,t)$ such that $D\bu/Dt =\bQ(\bx,\,t)$, we have 
\bel{ertel3}
\frac{D^2\bw}{Dt^2} = \bw\cdot\nabla\left(\frac{D\bu}{Dt}\right) = \bw\cdot\nabla\bQ\,.
\ee
In each of the cases (1-3) above $\bQ$ is readily identifiable and thus we have $\bdb$
\bel{a-dyn}
\frac{D\ba}{Dt} = \bw\cdot\nabla\bQ =: \bdb(\bx,\,t)\,,
\ee
thereby completing the quartet of vectors $(\bu,\,\bw,\,\ba,\,\bdb)$. In \S\ref{lagev} it will 
be shown that knowledge of the quartet of vectors $(\bu,\,\bw,\,\ba,\,\bdb)$ determines the 
quaternion-frame, which is a completely natural ortho-normal frame for the Lagrangian dynamics. 
Modulo a rotation around $\bw$, the quaternion-frame turns out to be the Frenet-frame attached 
to lines of constant $\bw$. Although usually credited 
to Ertel \cite{Er42}, the result in (\ref{ertel2}), which involves the cancellation of nonlinear 
terms of $O(|\bw||\nabla\bu|^2)$, actually goes much further back in the literature than this; 
see \cite{Ohk93,KZ,Trus,Viudez,Ohk95}. While Ertel's Theorem above enables us to find a $\bdb$ 
as in cases (1-3), $\bdb$ must be determined by other means in case (4). 

\subsection{Blow-up in the three-dimensional Euler equations}

The general picture of Lagrangian evolution and the associated quaternion frame is given in 
\S\ref{lagev}. Thereafter this paper will focus on the three-dimensional incompressible Euler 
equations (\ref{eul1}) (see \S\ref{Euler}) and the global existence of solutions (see 
\S\ref{BKMsect}). 

Many generations of mathematicians could testify to the deceptive simplicity of the Euler 
equations. The work of the late Victor Yudovich \cite{VY}, who 
proved the existence and uniqueness of weak solutions of the two-dimensional Euler equations 
with $\bom_{0}\in L^{\infty}$ on unbounded domains, will be remembered as a mile-stone in Euler 
dynamics.  In the three-dimensional case, while many special solutions are known in terms 
of simple functions \cite{GKB,Saffbk,MKO}, and powerful results have been found on weak and 
distributional solutions (see Majda \& Bertozzi \cite{MB01}) yet the fundamental problem 
of whether solutions exist for arbitrarily long times or become singular in a finite time 
still remains open. In physical terms, singular behaviour could potentially occur if a vortex 
is resolvable only by length scales decreasing to zero in a finite time.  While a review of 
certain aspects of the three-dimensional Euler singularity problem will form part of the later 
sections of this review, the regularity problem for the Navier-Stokes equations will not be 
considered; the interested reader should consult \cite{Lady,CFbk,FMRT}. 
\par\smallskip
In the first demonstrable case of Euler blow-up, Stuart \cite{jts1,jts2,jts3} considered 
solutions of the three-dimensional Euler equations that had linear dependence in two variables 
$x$ and $z$; the resulting differential equations in the remaining independent variables $y$ 
and $t$ displayed finite time singular behaviour. Stuart then showed how the method of 
characteristics leads to the construction of a complete class of singular solutions \cite{jts1}. 
This type of singularity has infinite energy because the solution is linearly stretched in the 
both the $x$ and $z$ directions.  In a similar fashion, Gibbon, Fokas \& Doering \cite{GFD} 
considered another class of infinite energy solutions whose third component of velocity is 
linear in $z$ so that the velocity field takes the form 
$\bu = \left\{u_{1}(x,y,t),\,u_{2}(x,y,t),\,z\gamma(x,y,t)\right\}$. These generalize the 
Burgers' vortex \cite{MKO} and represent tube and ring-like structures depending on the sign 
of $\gamma(x,y,t)$. Strong numerical evidence of singular behaviour on a periodic $x-y$ 
cross-section found by Ohkitani and Gibbon \cite{OG} was confirmed by an analytical proof of 
blow-up by Constantin \cite{IMRN}. Subsequently Gibbon, Moore and Stuart \cite{GMS} found two 
explicit singular solutions using the methods outlined in \cite{jts1}.
\par\smallskip
The Beale-Kato-Majda (BKM) theorem \cite{BKM} has been the main cornerstone of the analysis 
of potential \textit{finite energy} Euler singularities\,: one version of this theorem is 
that $\|\bom\|_{\infty}$ must satisfy (see \S\ref{BKMsect} for a more precise statement)
\bel{BKMintro}
\int_{0}^{T}\|\bom\|_{\infty}\,d\tau < \infty\,,
\ee
for a global solution to exist up to time $T$. The most important feature of (\ref{BKMintro}) 
is that it is single, simple criterion which is easily monitored. Several refinements of the 
BKM-Theorem exist in addition to those by Ponce \cite{Ponce}, who replaced $\|\bom\|_{\infty}$ 
by $\|S\|_{\infty}$, and the BMO-version proved by Kozono and Taniuchi \cite{KT}. In particular, 
these take account of the direction in which vorticity grows. The work of Constantin \cite{Const94},
and Constantin, Fefferman \& Majda \cite{CFM}, reviewed in \S\ref{CFMsub}, deserves specific 
mention. They were the first to make a precise mathematical formulation of how the misalignment 
of vortex lines might lead to, or prevent, a singularity. 
\S\ref{DHYsub} is devoted to a 
review of the work of Deng, Hou \& Yu \cite{DHY1,DHY2} who have established different criteria 
on vortex lines. In \S\ref{direc2}, quaternions are considered as an alternative way of looking 
at the direction of vorticity \cite{GHKR}, which provides us with a different direction of 
vorticity theorem based on the Hessian matrix of the pressure (\ref{Hessdef}). Further 
discussion and references are left to \S\ref{BKMsect}.

\subsection{Definition \& properties of quaternions}\label{quatdef}

In terms of any scalar $p$ and any 3-vector $\bq$, the quaternion $\bfq = [p,\,\bq]$ is defined 
as (Gothic fonts denote quaternions)
\bel{quatdef1}
\bfq = [p,\,\bq] = pI - \sum _{i=1}^{3}q_{i}\sigma_{i}\,,
\ee
where $\{\sigma_{1},\,\sigma_{2},\,\sigma_{3}\}$ are the three Pauli spin-matrices defined by
\bel{psm1}
\sigma_{1} = \left(\begin{array}{rr}
0 & i\\
i & 0
\end{array}\right)\,,
\hspace{2cm}
\sigma_{2} = \left(\begin{array}{rr}
0 & 1\\
-1 & 0
\end{array}\right)\,,
\hspace{2cm}
\sigma_{3} = \left(\begin{array}{rr}
i & 0\\
0 & -i
\end{array}\right)\,,
\ee
and which obey the relations $\sigma_{i}\sigma_{j} = -\delta_{ij}I-\epsilon_{ijk}\sigma_{k}$\,.
$I$ is the $2\times 2$ unit matrix. These relations then give a non-commutative multiplication 
rule
\bel{quatdef2}
\bfq_{1}\cast\bfq_{2} = [p_{1}p_{2} - \bq_{1}\cdot\bq_{2},\,p_{1}\bq_{2} + p_{2}\bq_{1} + 
\bq_{1}\times\bq_{2}]\,.
\ee
It can easily be demonstrated that quaternions are associative. One of the main properties of 
quaternions not shared by 3-vectors is the fact that they have an inverse; the inverse of 
$\bfq$ is $\bfq^{*} = [p,\,-\bq]$ which means that $\bfq\cast\bfq^{*} = [p^2 + q^2,\,0] = 
(p^2 + q^2)[1,\,0]$; of course, $[1,\,0]$ really denotes a scalar so if $p^2 + q^2 = 1$, 
$\bfq$ is a unit quaternion $\bhfq$. 
 
A quaternion of the type $\bfw = [0,\,\bw]$ is called a pure quaternion, with the product 
between two of them expressed as
\bel{pure1}
\bfw_{1}\cast\bfw_{2} = [0,\,\bw_{1}]\cast[0,\,\bw_{2}] = 
[-\bw_{1}\cdot\bw_{2},\,\bw_{1}\times\bw_{2}]\,.
\ee
In fact there is a quaternionic version of the gradient operator $\bnabla = [0,\,\nabla]$ 
which, when acting upon a pure quaternion $\mathfrak{u} = [0,\,\bu]$, gives
\bel{pure2}
\bnabla \cast\mathfrak{u} = [-\hbox{div}\,\bu,\,\hbox{curl}\,\bu]\,.
\ee
If the field $\bu$ is divergence-free, as for an incompressible fluid, then 
\bel{pure3}
\bnabla \cast\mathfrak{u} = [0,\,\bom]\,.
\ee
This pure quaternion incorporating the vorticity will be used freely in future sections.
\par\smallskip
It has been mentioned already in Section \ref{Hintro} that quaternions are used in the aerospace 
and computer animation industries to avoid difficulties with Euler angles. Here the relation is 
briefly sketched between quaternions and one of the many ways that have been used to describe 
rotating bodies in the rich and long-standing literature of classical mechanics -- for more 
see \cite{GH06a}. Whittaker \cite{Whitt1944}
shows how quaternions and the \textit{Cayley-Klein parameters} \cite{Klein04} are intimately 
related and gives explicit formulae relating these parameters to the Euler angles. Let $\bhfq = 
[p,\,\bq]$ be a unit quaternion with inverse $\bhfq^{*} = [p,\,-\bq]$ where $p^{2}+q^{2} = 1$. 
For a pure quaternion $\bfr = [0,\,\br]$ there exists a transformation from $\bfr \to \bfr' = 
[0,\,\br']$
\bel{r1}
\bfr' = \bhfq\cast\bfr\cast\bhfq^{*}\,.
\ee
This associative product can explicitly be written as
\bel{r2}
\bfr' = \bhfq\cast\bfr\cast\bhfq^{*} 
= [0,\, (p^{2}-q^{2})\br + 2p(\bq\times\br)+ 2\bq(\br\cdot\bq)]\,.
\ee
Choosing $p=\pm\cos\shalf\theta$ and $\bq = \pm\,\bhn\sin\shalf\theta$, where $\bhn$ 
is the unit normal to $\br$, we find that 
\bel{r3}
\bfr' = \bhfq\cast\bfr\cast\bhfq^{*}
= [0,\, \br\cos\theta + (\bhn\times\br) \sin\theta] \equiv O(\theta,\,\bhn)\br\,,
\ee
Equation (\ref{r3}) is the Euler-Rodrigues formula for the rotation $O(\theta,\,\bhn)$ 
by an angle $\theta$ of the vector $\br$ about its unit normal $\bhn$\,; $\theta$ and $\bhn$ 
are called the Euler parameters. With the choice of $p$ and $\bq$ above $\bhfq$ is given by 
\bel{r4}
\bhfq = \pm [\cos\shalf\theta,\,\bhn\sin\shalf\theta]\,.
\ee
The elements of the unit quaternion $\bhfq$ are the Cayley-Klein parameters which are related to
the Euler angles and which form a representation of the Lie group $SU(2)$. All terms  in the 
(\ref{r2}) are quadratic in $p$ and $\bq$, and thus possess the well-known $\pm$ equivalence 
which is an expression of the fact that $SU(2)$ covers $SO(3)$ twice.
\par\smallskip
To investigate the  map (\ref{r1}) when $\bhfp$ is time-dependent, the Euler-Rodrigues formula 
in (\ref{r3}) can be written as 
\bel{rot2}
\bfr'(t) = \bhfp\cast\bfr\cast\bhfp^{*}
\hspace{1cm}\Rightarrow\hspace{1cm}
\bfr = \bhfp^{*}\cast\bfr'(t)\cast\bhfp\,.
\ee
Thus $\dot{\bfr}'$ has a time derivative given by
\beq\label{rot3}
\dot{\bfr}'(t) &=& \dot{\bhfp}\cast(\bhfp^{*}\cast\bfr'\cast\bhfp)\cast\bhfp^{*}
+ \bhfp\cast(\bhfp^{*}\cast\bfr'\cast\bhfp)\cast\dot{\bhfp}^{*}\nonumber\\
&=& \dot{\bhfp}\cast\bhfp^{*}\cast\bfr' + \bfr'\cast\bhfp\cast\dot{\bhfp}^{*}\nonumber\\
&=& (\dot{\bhfp}\cast\bhfp^{*})\cast\bfr' + \bfr'\cast(\dot{\bhfp}\cast\bhfp^{*})^{*}\nonumber\\
&=& (\dot{\bhfp}\cast\bhfp^{*})\cast\bfr' - ((\dot{\bhfp}\cast\bhfp^{*})\cast\bfr')^{*}\,,
\eeq
having used the fact on the last line that because $\bfr'$ is a pure quaternion, $\bfr'^{*} 
= -\bfr'$.
Because $\bhfp = [p,\,\bq]$ is of unit length, and thus $p\dot{p} + q\dot{q} = 0$, 
this means that $\dot{\bhfp}\cast\bhfp^{*}$ is also a pure quaternion 
\bel{rot4}
\dot{\bhfp}\cast\bhfp^{*} = [0,\,\shalf\bcapom_{0}(t)]\,.
\ee
The 3-vector entry in (\ref{rot4}) defines the angular frequency $\bcapom_{0}(t)$ as 
$\bcapom_{0} = 2(-\dot{p}\bq +\dot{\bq}p - \dot{\bq}\times\bq)$ thereby giving the 
well-known formula for the rotation of a rigid body
\bel{rot5}
\dot{\br}' = \bcapom_{0}\times\br'\,.
\ee 
For a Lagrangian particle, the equivalent of $\bcapom_{0}$ is the Darboux vector 
$\bD_{a}$ in Theorem \ref{abthm} of \S\ref{lagev}. This Theorem is the main result 
of this paper and is the equivalent of (\ref{rot5}) for a Lagrangian particle 
undergoing rotation in flight.
\par\smallskip
Finally, it can easily be seen that Hamilton's relation in terms of hyper-complex numbers 
$i^{2}=j^{2}=k^{2}= ijk = -1$ will generate the rule in (\ref{quatdef2}) if $\bfq$ is written 
as a 4-vector $\bfq = p + i q_{1} + j q_{2} + k q_{3}$. Sudbery's paper is still the best 
source for a study of the functional properties of quaternions \cite{Sudbery79}; he discusses 
how various results familiar for functions over a complex field, such as the Cauchy-Riemann 
equations, Cauchy's Theorem and integral formula, together with the Laurent expansion (but not 
conformal mappings) have their parallels for quaternionic functions. More recent work on further 
analytical properties can be found in \cite{Fokasquat}. 

\section{\large Lagrangian evolution equations and an ortho-normal frame}\label{lagev}

This section sets up the mathematical foundation concerning the association of quaternion frames 
and can be found in the paper by Gibbon and Holm \cite{GH06a}. Let us repeat the Lagrangian 
evolution equations for a vector field $\bw$ satisfying (\ref{w-dyn1}) and (\ref{a-dyn2})
\bel{aw-dyn}
\frac{D\bw}{Dt} = \ba(\bx,\,t)\,,
\hspace{3cm}
\frac{D\ba}{Dt} = \bdb(\bx,\,t)\,.
\ee
\par\vspace{-3mm}\noindent
\bc
\begin{minipage}[c]{.75\textwidth}
\begin{pspicture}
\psframe(0,0)(5,5)
\thicklines
\qbezier(0,1)(4,2.5)(0,4)
\thinlines
\put(2.01,2.43){\makebox(0,0)[b]{$\bullet$}}
\put(1.4,2.5){\makebox(0,0)[b]{\scriptsize$(\bx_{1},t_{1})$}}
\thinlines
\put(2,2.5){\vector(0,1){1}}
\put(2,3.7){\makebox(0,0)[b]{$\bhw$}}
\put(2,2.5){\vector(-2,-1){1}}
\put(.7,1.8){\makebox(0,0)[b]{$\bhchi_{a}$}}
\put(2,2.5){\vector(1,0){1}}
\put(3.8,2.4){\makebox(0,0)[b]{$\bhw\times\bhchi_{a}$}}
\thicklines
\qbezier(7,1)(6,2.5)(8,4)
\thinlines
\put(6.77,2.45){\makebox(0,0)[b]{$\bullet$}}
\put(6.2,2.5){\makebox(0,0)[b]{\scriptsize$(\bx_{2},t_{2})$}}
\thicklines
\qbezier[50](2,2.5)(3,.5)(6.7,2.5)
\thinlines
\put(6.73,2.5){\vector(1,4){.3}}
\put(7,3.8){\makebox(0,0)[b]{$\bhw$}}
\put(6.7,2.5){\vector(4,-1){1}}
\put(8.4,2.8){\makebox(0,0)[b]{$\bhw\times\bhchi_{a}$}}
\put(6.7,2.5){\vector(4,1){1}}
\put(8.1,2.1){\makebox(0,0)[b]{$\bhchi_{a}$}}
\put(3,1.2){\vector(1,0){.6}}
\put(4,.7){\makebox(0,0)[b]{\small tracer particle trajectory}}
\put(4.5,1.4){\vector(4,1){.6}}
\thinlines
\end{pspicture}
\end{minipage}
\ec
\bc
\vspace{-1mm}
\begin{minipage}[r]{\textwidth}
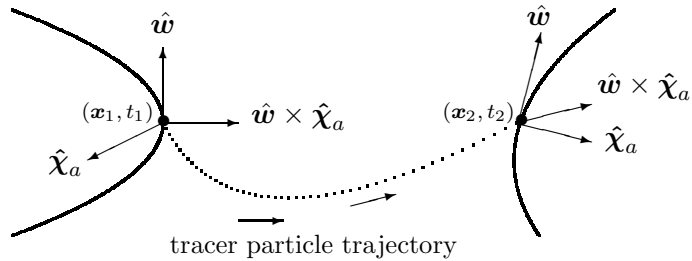
\textbf{Figure 1:} {\small The dotted line represents the tracer particle $(\bullet)$ path  
moving from $(\bx_{1},t_{1})$ to $(\bx_{2},t_{2})$. The solid curves represent lines of constant 
$\bw$ to which $\bhw$ is a unit tangent vector. The  orientation of the quaternion-frame
$(\bhw,\,\bhchi_{a},\,\bhw\times\bhchi_{a})$ is  shown at the two space-time points; note that 
this is not the Frenet-frame corresponding to the particle path but to lines of constant $\bw$.}
\end{minipage}
\ec
\par\smallskip
Given the Lagrangian equations in (\ref{aw-dyn}), define the scalar $\alpha_{a}$ and the 
3-vector $\bchi_{a}$ as\footnote{The role of null points $\bw = 0$ is not yet clear although, 
as \S\ref{Euler} shows, this problem is neatly avoided by the Euler fluid equations.  It has 
been discussed at greater length in \cite{GH06a}.}
\bel{la2a}
\alpha_{a} = |\bw|^{-1}(\bhw\cdot\ba)\,,\hspace{1.5cm}\bchi_{a} = |\bw|^{-1}(\bhw\times\ba)\,,
\hspace{1.5cm}
w \neq 0\,.
\ee
Moreover, let $\alpha_{b}$ and $\bchi_{b}$ be defined as in (\ref{la2a}) for $\alpha_{a}$ and 
$\bchi_{a}$ with $\ba$ replaced by $\bdb$. The 3-vector $\ba$ can be decomposed into parts 
that are parallel and perpendicular to $\bw$ (and likewise the same for $\bdb$)
\bel{decom1}
\ba = \alpha_{a}\bw + \bchi_{a}\times\bw 
= [\alpha_{a},\,\bchi_{a}]\cast[0,\,\bw]\,,
\ee
and thus the quaternionic product is summoned in a natural manner. By definition, the 
growth rate $\alpha_{a}$  of the magnitude $|\bw|$ obeys 
\bel{la3}
\frac{D|\bw|}{Dt} = \alpha_{a}|\bw|\,,
\ee
while the unit tangent vector $\bhw = \bw w^{-1}$ satisfies 
\bel{la5}
\frac{D\bhw}{Dt} = \bchi_{a}\times \bhw\,.
\ee
Now identify the quaternions\footnote{Dropping the $a\,,b$ labels and normalizing, the 
Cayley-Klein parameters are $\bhfq = [\alpha,\,\bchi](\alpha^2 + \chi^2)^{-1/2}$.}
\bel{ls6}
\bfq_{a} 
= [\alpha_{a},\,\bchi_{a}]\,,\hspace{2cm}\bfq_{b} 
= [\alpha_{b},\,\bchi_{b}]\,,
\ee
and let $\bfw = [0,\,\bw]$ be the pure quaternion satisfying  the Lagrangian evolution equation 
(\ref{aw-dyn}) with $\bfq_{a}$ defined in (\ref{ls6}).  Then the first equation in (\ref{aw-dyn}) 
can automatically be re-written equivalently in the quaternion form
\bel{lem1}
\frac{D\bfw}{Dt} 
= [0,\,\ba] 
= [0,\, \alpha_{a}\bw + \bchi_{a}\times\bw ] 
= \bfq_{a}\cast\bfw\,.
\ee
Moreover, if $\ba$ is differentiable in the Lagrangian sense as in
(\ref{aw-dyn}) then it  is clear that a similar decomposition for $\bdb$ as
that for $\ba$ in (\ref{decom1}) gives 
\bel{la9}
\frac{D^{2}\bfw}{Dt^{2}} 
= [0,\,\bdb] 
= [0,\, \alpha_{b}\bw + \bchi_{b}\times\bw ] 
= \bfq_{b}\cast\bfw \,.
\ee
Using the associativity property, compatibility of (\ref{la9}) and (\ref{lem1}) implies 
that ($|\bw| \neq 0$)
\bel{la10}
\left(\frac{D\bfq_{a}}{Dt} + \bfq_{a}\cast\bfq_{a} 
-\bfq_{b}\right)\cast\bfw = 0\,,
\ee
which establishes a \textit{Riccati relation} between $\bfq_{a}$ and
$\bfq_{b}$
\bel{Ric1}
\frac{D\bfq_{a}}{Dt} + \bfq_{a}\cast\bfq_{a} = \bfq_{b}\,.
\ee
This relation is closely allied to the ortho-normal quaternion-frame\footnote{According to 
Hanson \cite{Ha2006} the the quaternion-frame is similar to the Bishop-frame in computer 
graphics.} $(\bhw,\,\bhchi_{a},\,\bhw\times\bhchi_{a})$ whose equations of 
motion are given as follows\,:
\begin{theorem}\label{abthm}\cite{GH06a} 
The ortho-normal quaternion-frame $(\bhw,\,\bhchi_{a},\,\bhw\times\bhchi_{a})\in SO(3)$ 
has Lagrangian time derivatives expressed as
\beq\label{abframe3}
\frac{D\bhw}{Dt}&=& \bD_{ab}\times\bhw\,,\\
\frac{D(\bhw\times\bhchi_{a})}{Dt} &=& \bD_{ab}\times(\bhw\times\bhchi_{a})\,,\label{abframe4}
\\
\frac{D\bhchi_{a}}{Dt} &=& \bD_{ab}\times\bhchi_{a}\,,\label{abframe5}
\eeq
where the Darboux angular velocity vector $\bD_{ab}$ is defined as
\bel{abframe6}
\bD_{ab} = \bchi_{a} + \frac{c_{b}}{\chi_{a}}\bhw\,,\hspace{2cm}
c_{b} = \bhw\cdot(\bhchi_{a}\times\bchi_{b})\,.
\ee
\end{theorem}
\par\smallskip\noindent
\textbf{Remark:} The analogy with the formula for a rigid body is obvious when compared to (\ref{rot5}).
but the Darboux angular velocity vector $\bD_{ab}$ is itself a function of $\bchi\,,~\bhw$ and other
variables and sits in a two-dimensional plane. In turn this is driven by 
$c_{b} = \bhw\cdot(\bhchi_{a}\times\bchi_{b})$
for which $\bdb$ must be known. Given this it may then possible to numerically solve equations 
(\ref{abframe3}) -- (\ref{abframe6}) for the particle paths.
\par\smallskip\noindent
\textbf{Proof\,:} To find an expression for the Lagrangian time derivatives of the components of 
the frame  $(\bhw,\,\bhchi_{a},\,\bhw\times\bhchi_{a})$ requires the derivative of $\bhchi_{a}$. 
To find this it is necessary to use the fact that the 3-vector $\bdb$ can be expressed in this 
ortho-normal frame as the linear combination
\bel{b1}
w^{-1}\bdb = \alpha_{b}\,\bhw + c_{b}\bhchi_{a} + d_{b}(\bhw\times\bhchi_{a})\,.
\ee
where $c_{b}$ is defined in (\ref{abframe6}) and $d_{b} = -\,(\bhchi_{a}\cdot\bchi_{b})$.
The 3-vector product $\bchi_{b} = w^{-1}(\bhw\times\bdb)$ yields 
\bel{bchibdef}
\bchi_{b} = c_{b}(\bhw\times\bhchi_{a}) - d_{b}\bhchi_{a}\,.
\ee
To find the Lagrangian time derivative of $\bhchi_{a}$, we use the 3-vector part of the 
equation for the quaternion $\bfq_{a} = [\alpha_{a},\,\bchi_{a}]$ in Theorem \ref{abthm}
\bel{abframe1}
\frac{D\bchi_{a}}{Dt} = - 2\alpha_{a}\bchi_{a} + \bchi_{b}\,,
\hspace{1cm}
\Rightarrow
\hspace{1cm}
\frac{D\chi_{a}}{Dt} = -2\alpha_{a}\chi_{a} - d_{b}\,,
\ee
where $\chi_{a} = |\bchi_{a}|$. Using (\ref{bchibdef}) and (\ref{abframe1}) there follows
\bel{abframe2}
\frac{D\bhchi_{a}}{Dt} = c_{b}\chi_{a}^{-1}(\bhw\times\bhchi_{a})\,,
\hspace{2cm}
\frac{D(\bhw\times\bhchi_{a})}{Dt} = \chi_{a}\,\bhw - c_{b}\chi_{a}^{-1}\bhchi_{a}\,,
\ee
which gives equations (\ref{abframe3})-(\ref{abframe6}).\hspace{9cm}$\blacksquare$
\par\medskip
How to find the rate of change of acceleration represented by the $\bdb$-vector is an 
important question regarding computing the paths of passive tracer particles where
$\bdb$ is not known through Ertel's Theorem.  The result that follows describes the 
evolution of $\bfq_{b}$ in terms of three arbitrary scalars. 
\begin{theorem}\label{b-field}
\cite{GH06a} The Lagrangian time derivative of $\bfq_{b}$ can be expressed as 
\bel{qbev1}
\frac{D\bfq_{b}}{Dt} = \bfq_{a}\cast\bfq_{b} + 
\lambda_{1}\bfq_{b} + \lambda_{2}\bfq_{a}  + \lambda_{3}\mathbb{I}\,,
\ee
where the $\lambda_{i}(\bx,\,t)$ are arbitrary scalars $(\mathbb{I} = [1,\,0])$.
\end{theorem}
\par\medskip\noindent
\textbf{Proof\,:} 
To establish (\ref{qbev1}), we differentiate the orthogonality relation $\bchi_{b}\cdot\bhw = 0$ 
and use the Lagrangian derivative of $\bhw$
\bel{press2}
\frac{D\bchi_{b}}{Dt} = \bchi_{a}\times\bchi_{b} + \bs_{0}\,,
\hspace{1cm}
\hbox{where}
\hspace{1cm}
\bs_{0} = \mu\bchi_{a} +  \lambda\,\bchi_{b}\,.
\ee
$\bs_{0}$ lies in the plane perpendicular to $\bhw$ in which $\bchi_{a}$ and $\bchi_{b}$ 
also lie and $\mu =\mu(\bx,t)$ and $\lambda = \lambda(\bx,t)$ are arbitrary scalars. 
Explicitly differentiating $\bchi_{b} = w^{-1}(\bhw\times\bdb)$ gives
\bel{alph1}
w^{-1}\bhw\left(\bchi_{a}\cdot\bdb\right) + \bs_{0} = 
- \alpha_{a}\bchi_{b} - \alpha_{b}\bchi_{a} + w^{-1}\bhw\left(\bchi_{a}\cdot\bdb\right) + 
w^{-1}\left(\bhw\times\frac{D\bdb}{Dt}\right)\,,
\ee
which can easily be manipulated into
\bel{alph2}
\bhw\times\left\{\frac{D\bdb}{Dt} - \alpha_{b}\,\ba  - \alpha_{a}\,\bdb\right\} = w\,\bs_{0}\,.
\ee
This means that
\bel{alph3}
\frac{D\bdb}{Dt} = \alpha_{b}\ba + \alpha_{a}\bdb + \bs_{0} \times\bw + \varepsilon\bw\,,
\ee
where $\varepsilon = \varepsilon(\bx,t)$ is a third unknown scalar in addition to $\mu$ 
and $\lambda$ in (\ref{press2}). Thus the Lagrangian derivative of 
$\alpha_{b} = w^{-1}(\bhw\cdot\bdb)$ is
\bel{alph4}
\frac{D\alpha_{b}}{Dt} = \alpha\alpha_{b} + \bchi_{a}\cdot\bchi_{b} + \varepsilon\,.
\ee
Lagrangian differential relations have now been found for $\bchi_{b}$ and $\alpha_{b}$, but at 
the price of introducing the triplet of unknown coefficients $\mu,~\lambda$, and $\varepsilon$ 
which are re-defined as
\bel{tetrad1}
\lambda =\alpha_{a} + \lambda_{1}\,,\hspace{1cm}
\mu = \alpha_{b} + \lambda_{2}\,,\hspace{1cm}
\varepsilon = -2\bchi_{a}\cdot\bchi_{b} + \lambda_{2}\alpha_{a} + \lambda_{1}\alpha_{b}+ \lambda_{3}\,.
\ee
The new triplet has been subsumed into (\ref{qbev1}). Then (\ref{press2}) 
and (\ref{alph4}) can be written in the quaternion form (\ref{qbev1}).\hspace{6cm}$\blacksquare$

\section{\large Quaternions and the incompressible $3D$ Euler equations}\label{Euler}

The results of the previous section on Lagrangian flows are immediately applicable to 
the incompressible Euler equations, but to present them in this manner is actually to 
do so in the chronologically reverse order in which they were first developed. Looking 
ahead in this section, the variables $\alpha$ and $\bchi$ in (\ref{triad5a}) for the 
Euler equations, and the two coupled differential equations that they satisfy (\ref{alchi}), 
were first written down almost ten years ago in \cite{GGH,GGK} without the help of 
quaternions. It was then discovered in \cite{Gibbon02} that these equations could be 
combined to form a quaternionic Riccati equation. Finally, the more recent paper 
\cite{GHKR}, in combination with \cite{GH06a}, put all these results in the form 
expounded in this present paper. Because data for the three-dimensional Euler equations 
gets very rough very quickly it should be understood that all our manipulations are formal.
\par\smallskip
In \S\ref{lagev} it was shown that a knowledge of the quartet of vectors 
$(\bu,\,\bw,\,\ba,\,\bdb)$ is necessary to be able to use the results of 
Theorem \ref{abthm}. With $\bw\equiv\bom$ and $\bom = \hbox{curl}\,\bu$ the vortex 
stretching vector is $\ba = \bom\cdot\nabla\bu$. Thus the $\bw$- and $\bu$-fields 
are not independent in this case. Within $\ba = \bom\cdot\nabla\bu$, the dot-product 
of $\bom$ sees only the symmetric part of the velocity gradient matrix $\nabla\bu$, 
which is the strain matrix $S_{ij} = \shalf\left(u_{i,j} + u_{j,i}\right)$ defined 
in (\ref{straindef}).  With $\ba = \bom\cdot\nabla\bu = S\bom$, the triad of vectors 
is 
\bel{triad3}
(\bu,\,\bw,\,\ba) \equiv (\bu,\,\bom,\,S\bom)\,.
\ee
To find the $\bdb$-field, Ertel's Theorem of \S\ref{applic} comes to the rescue. The $D\bu/Dt$ 
within the right hand side of (\ref{eul1}) (with $\bw=\bom$) obeys Euler's equation $D\bu/Dt = 
-\nabla p$, so we have 
\bel{ertel1}
\bdb = \bom\cdot\nabla\left(\frac{D\bu}{Dt}\right) = - P\bom\,,
\ee
where $P$ is the Hessian of the pressure defined in (\ref{Hessdef}). The quartet of vectors 
necessary to make Theorem \ref{abthm} work is now in place
\bel{triad4}
(\bu,\,\bw,\,\ba,\,\bdb) \equiv (\bu,\,\bom,\,S\bom,\,-P\bom)\,.
\ee
The table below discusses three quartets $(\bu,\,\bw,\,\ba,\,\bdb)$ for the Euler fluid 
equations\,:
$$
\begin{array}{c|ccc|c}
\quad\bu\quad & \quad\bw\quad & \quad\ba\quad & \quad\bdb\quad & \hbox{\small Material~Deriv}\\\hline
\hbox{\small Euler} & \quad\bx\quad & \quad\bu\quad & \quad-\nabla p\quad & (\ref{mat1})\\
\hbox{\small Euler} & \quad\bu\quad & \quad-\nabla p\quad & \quad?\quad& (\ref{mat1})\\
\hbox{\small Euler} & \quad \bom\quad & \quad S\bom \quad & \quad -P\bom\quad& (\ref{mat1})
\end{array}
$$
{\small Table 1\,: The entries above are three of the possibilities for finding a $\bdb$-field 
given the triplet $(\bu,\,\bw,\,\ba)$. The third line is the result (\ref{triad4}) while $\bdb$ 
is unknown for the second line.}
\par\medskip
Using the definitions in \S\ref{lagev} the scalar $\alpha$ and the 3-vector $\bchi$ are 
defined as
\bel{triad5a}
\alpha = \bhom\cdot S\bhom\,,\hspace{2cm}\bchi = \bhom\times S\bhom\,,
\ee
together with the definitions for $\alpha_{p}$ and $\bchi_{p}$
\bel{triad5b}
\alpha_{p} = \bhom\cdot P\bhom\,,\hspace{2cm}\bchi_{p} = \bhom\times P\bhom\,.
\ee
$\alpha$ in (\ref{triad5a}) is now identified as the same as that in Constantin 
\cite{Const94} who has expressed it as an explicit Biot-Savart formula\footnote{Everywhere 
in \cite{Const94,CFM,DC3,DC6} the unit vector of vorticity is 
designated as $\xi$ whereas here we use $\bhom$.}. $\ba = S\bom$ can be decomposed 
into parts that are parallel and perpendicular to $\bom$ 
\bel{triad6}
S\bom = \alpha\bom + \bchi\times\bom = [\alpha,\,\bchi]\cast[0,\,\bom]\,.
\ee
By definition, the growth rate $\alpha$ of the scalar magnitude $|\bom|$ 
and the unit tangent vector $\bhom$ obey
\bel{triad7}
\frac{D|\bom|}{Dt} = \alpha |\bom|\,,
\hspace{3cm}
\frac{D\bhom}{Dt} = \bchi\times \bhom\,,
\ee
which show that  $\alpha$ drives the growth or collapse of vorticity and 
$\bchi$ determines the rate of swing of $\bhom$ around $S\bom$. 
Now identify the quaternions 
\bel{triad8}
\bfq = [\alpha,\,\bchi]\,,\hspace{2cm}\bfq_{p} = [\alpha_{p},\,\bchi_{p}]\,.
\ee
The equivalent of the Riccati equation (\ref{Ric1}) is\footnote{In principle (\ref{Rictriad}) 
can be linearized to a zero-eigenvalue Schr\"odinger equation in quaternion form with $\bfq_{p}$ 
as the potential, although it is not clear how to proceed from that point.}
\bel{Rictriad}
\frac{D\bfq}{Dt} + \bfq\cast\bfq + \bfq_{p} = 0\,,
\ee
which, when written explicitly in terms of $\alpha$--$\bchi$, becomes
\bel{alchi}
\frac{D\alpha}{Dt} + \alpha^2 - \chi^2 +\alpha_{p} = 0\,.
\hspace{2cm}
\frac{D\bchi}{Dt} + 2\alpha\bchi + \bchi_{p} = 0\,.
\ee
In Theorem \ref{abthm} we need to use $\bdb = -P\bom$ to calculate the path of 
the ortho-normal quaternion-frame $(\bhom,\,\bhchi,\,\bhom\times\bhchi)$. 
Specifically we must solve
\beq\label{abframe7}
\frac{D\bhom}{Dt}&=& \bD\times\bhom\,,\\
\frac{D(\bhom\times\bhchi)}{Dt} &=& \bD\times(\bhom\times\bhchi)\,,\label{abframe8}
\\
\frac{D\bhchi}{Dt} &=& \bD\times\bhchi\,,\label{abframe9}
\eeq
where the Darboux angular velocity vector $\bD$ is defined as
\bel{abframe10}
\bD = \bchi + \frac{c_{p}}{\chi}\bhom\,,\hspace{2cm}
c_{p} = -\bhom\cdot(\bhchi\times\bchi_{p})\,,
\ee
The pressure Hessian contributes to the angular velocity $\bD$ through the scalar coefficient 
$c_{p}$. To compute the fluid particle paths one would need data on the pressure Hessian $P$ 
as well as the vorticity $\bom$ and the strain matrix $S$.  It is here where the fundamental 
difference between the Euler equations and a passive problem is made explicit.  For the Euler 
equations the $\bdb$-field containing $P$ is not independent of $\bw \equiv\bom$ but is 
connected subtly and non-locally through the elliptic equation for the pressure (\ref{Trace}) 
which we repeat here
\bel{ellip1}
-Tr\,P = Tr (S^{2})-\shalf\omega^{2}\,.
\ee
\par\vspace{-3mm}\noindent
\bc
\begin{minipage}[c]{.75\textwidth}
\begin{pspicture}
\psframe(0,0)(5,5)
\thicklines
\qbezier(0,1)(4,2.5)(0,4)
\thinlines
\put(2.01,2.43){\makebox(0,0)[b]{$\bullet$}}
\put(1.4,2.5){\makebox(0,0)[b]{\scriptsize$(\bx_{1},t_{1})$}}
\thinlines
\put(2,2.5){\vector(0,1){1}}
\put(2,3.7){\makebox(0,0)[b]{$\bhom$}}
\put(2,2.5){\vector(-2,-1){1}}
\put(.7,1.8){\makebox(0,0)[b]{$\bhchi$}}
\put(2,2.5){\vector(1,0){1}}
\put(3.8,2.4){\makebox(0,0)[b]{$\bhom\times\bhchi$}}
\thicklines
\qbezier(7,1)(6,2.5)(8,4)
\thinlines
\put(6.77,2.45){\makebox(0,0)[b]{$\bullet$}}
\put(6.2,2.5){\makebox(0,0)[b]{\scriptsize$(\bx_{2},t_{2})$}}
\thicklines
\qbezier[50](2,2.5)(3,.5)(6.7,2.5)
\thinlines
\put(6.73,2.5){\vector(1,4){.3}}
\put(7,3.8){\makebox(0,0)[b]{$\bhom$}}
\put(6.7,2.5){\vector(4,-1){1}}
\put(8.4,2.8){\makebox(0,0)[b]{$\bhom\times\bhchi$}}
\put(6.7,2.5){\vector(4,1){1}}
\put(8.1,2.1){\makebox(0,0)[b]{$\bhchi$}}
\put(3,1.2){\vector(1,0){.6}}
\put(4,.7){\makebox(0,0)[b]{\small fluid particle trajectory}}
\put(4.5,1.4){\vector(4,1){.6}}
\thinlines
\end{pspicture}
\end{minipage}
\ec
\bc
\vspace{-1mm}
\begin{minipage}[r]{\textwidth}
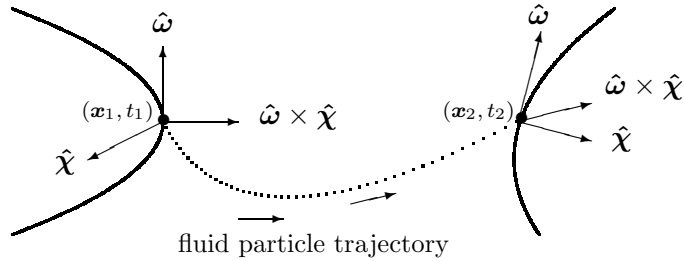
\textbf{Figure 2:} {\small The equivalent of Figure 1 but for the Euler equations with the 
dotted line representing an Euler fluid particle $(\bullet)$ path  moving from $(\bx_{1},t_{1})$ 
to $(\bx_{2},t_{2})$. The solid curves represent vortex lines to which $\bhom$ is a unit tangent 
vector. The  orientation of the quaternion-frame $(\bhom,\bhchi,~\bhom\times\bhchi)$ is  shown 
at the two space-time points; note that this is not the Frenet-frame corresponding to the particle 
path.}
\end{minipage}
\ec
\par\medskip
Theorem \ref{b-field} expresses the evolution of $\bfq_{p}$
\bel{qp1}
\frac{D\bfq_{p}}{Dt} = \bfq\cast\bfq_{p} + \lambda_{1}\bfq_{p} - \lambda_{2}\bfq  - 
\lambda_{3}\mathbb{I}\,,
\ee
in terms of the arbitrary scalars $\lambda_{i}(\bx,\,t)$. How these can be determined or handled in
terms of the incompressibility condition is not clear.

\section{\large The BKM Theorem \& the direction of vorticity}\label{BKMsect}

Three-dimensional Euler data becomes very rough very quickly\,; thus, understanding how vorticity grows
and in what direction, are fundamental questions that have yet to be definitively answered. Clearly
the vortex stretching term $\bom\cdot\nabla\bu = S\bom$, and the alignment of $\bom$ with the 
eigenvectors $\be_{i}$ of $S$, play a fundamental role in determining whether or not a singularity 
forms in finite time. Major
computational studies of this phenomenon can be found in Brachet \etal \cite{BMONMF,BMVPS}; Pumir \&
Siggia \cite{PS}; Kerr \cite{Kerr93,Kerr05}; Grauer \etal \cite{GMG}, Boratav \& Pelz \cite{BorPelz94},
Pelz \cite{Pelz01}, and Hou \& Li \cite{HL}. Studies of singularities in the complex time domain
of the two-dimensional Euler equations can be found in Pauls, Matsumoto, Frisch \& Bec \cite{PMFB} 
where an extensive literature is cited.
\par\smallskip
The BKM-theorem \cite{BKM} is the key result in studying the growth of Euler vorticity and 
possible singular behaviour. The domain $\mathbb{D}\subset\mathbb{R}^{3}$ in Theorem 
\ref{BKMthm} is taken to be a three-dimensional periodic domain for present purposes, which 
guarantees local existence in time of classical solutions (Kato \cite{Kato72}), although it 
is applicable on more general domains than this. One version is ($H^s$ denotes the standard 
Sobolev space):
\begin{theorem}\label{BKMthm} (Beale, Kato and Majda \cite{BKM}){\bf\,:}
On the domain $\mathbb{D} = [0,L]^{3}_{per}$ there exists a global solution of the Euler 
equations, $\bu \in C([0,\,\infty];H^{s})\cap C^{1}([0,\,\infty];H^{s-1})$ for $s\geq 3$ 
if, for every $T>0$
\bel{BKMthm1}
\int_{0}^{T}\|\bom\|_{\lin}\,d\tau < \infty\,.
\ee
\end{theorem}
The result can be stated the opposite way which is that no singularity can form at $T$ without 
$\int_{0}^{T}\|\bom\|_{\lin}\,d\tau = \infty$. Theorem \ref{BKMthm} has direct computational 
consequences. In a hypothetical computational experiment if one finds vorticity growth 
$\|\bom\|_{\lin} \sim (T - t)^{-\gamma}$ for some $\gamma > 0$, then the theorem says that 
$\gamma$ must satisfy $\gamma \geq 1$ for the observed singular behaviour to be real and not an 
artefact of the numerical computations. The reason is that if $\gamma$ is found to lie in 
the range $0< \gamma < 1$ then $\|\bom\|_{\lin}$ blows up whereas its time integral does not, 
thus violating the theorem. Of the many numerical calculations performed on Euler that by Kerr 
\cite{Kerr93,Kerr05}, using anti-parallel vortex tubes as initial data, was the first to see 
$\gamma$ pass the threshold with a critical value of $\gamma = 1$, followed by Grauer \etal 
\cite{GMG}, Boratav \& Pelz \cite{BorPelz94} and Pelz \cite{Pelz01}. Recent numerical calculations 
by Hou \& Li \cite{HL}, however, have contradicted the existence of a singularity: see 
\cite{Kerrhis} for a response and a discussion of the issues.  To fully settle this question 
will require more refined computations in tandem with analysis to understand the role played 
by the direction of vorticity growth. As indicated in \S1, the work of Constantin, Fefferman 
\& Majda \cite{CFM} (see also Constantin \cite{Const94}) was the first to make a precise 
mathematical formulation of how smooth the direction of vortex lines have to be that might 
lead to, or prevent, a singularity. \S\ref{CFMsub} is devoted to a short review of this work. 
Further papers by Cordoba \& Fefferman \cite{CF}, Deng, Hou \& Yu \cite{DHY1,DHY2} and Chae 
\cite{DC3,DC6} are variations on this theme. This approach, pioneered in \cite{CFM}, lays the 
mathematical foundation for the next generation of computational experiments, after the manner 
of Kerr \cite{Kerr93,Kerr05,Kerrhis} and Hou \& Li \cite{HL}, to check whether a singularity 
develops. \S\ref{DHYsub} is devoted to a description of the results in the papers by Deng, Hou 
\& Yu \cite{DHY1,DHY2} who have established different criteria on vortex lines. \S\ref{direc2} 
is devoted to an alternative direction of vorticity theorem proved in \cite{GHKR} based on the 
quaternion formulation of this paper.  
\par\smallskip
References and a more global perspective on the Euler equations can be found in the book 
by Majda and Bertozzi \cite{MB01}. Shnirelman \cite{Shnirel97} has constructed very weak 
solutions which have some realistic features but whose kinetic energy monotonically decreases 
in time and which are everywhere discontinuous and unbounded and for its dynamics in the more 
exotic function spaces see the papers by Tadmor \cite{Tad1} and Chae \cite{DC2,DC5,DC4}.

\subsection{The work of Constantin, Fefferman \& Majda}\label{CFMsub}

The obvious question regarding the BKM-criterion is whether the $L^\infty$-norm 
can be weakened to $L^p$ for $1\leq p \leq \infty$. This question was addressed 
by Constantin \cite{Const94} who placed further assumptions on the local nature 
of the vorticity and velocity fields. Consider the velocity field
\bel{C94a}
U_{1}(t) := \sup_{\bx}|\bu(\bx,\,t)|,
\ee
and the $L^{1}_{loc}$-norm of $\bom$ defined by
\bel{C94b}
\|\bom\|_{1,\,loc} = L^{-3}\sup_{x}\int_{|\by|\leq L}|\bom(\bx + \by)|d^{3}y\,,
\ee
where $L$ is some outer length scale in the Euler flow which could be taken to be unity.
Now assume that the unit vector of vorticity is Lipschitz
\bel{C94c}
|\bhom(\bx,\,t) - \bhom(\by,\,t)| \leq \frac{|\bx - \by|}{\rho_{0}(t)}
\ee
for $|\bx - \by| \leq L$ and for some length $\rho_{0}(t)$. Then the following result is stated 
in Constantin \cite{Const94} and re-stated and proved in Constantin, Fefferman \& Majda \cite{CFM}:
\begin{theorem}\label{CFMthm2} (Constantin \cite{Const94}; Constantin, Fefferman \& 
Majda \cite{CFM}){\bf\,:} Assume that the initial vorticity $\bom_{0}$ is smooth and compactly 
supported and assume that a solution of the Euler equations satisfies
\bel{C94d}
\int_{0}^{T}\|\bom(\cdot\,,\,s)\|_{1,\,loc} \left(\frac{L}{\rho_{0}(s)}\right)^{3}ds < \infty\,,
\hspace{2cm}
\int_{0}^{T}\frac{U(s)}{\rho_{0}(s)}\,ds < \infty\,.
\ee
Then 
\bel{C94e}
\sup_{0\leq t\leq T} \frac{\|\bom(\cdot\,,\,t)\|_{\infty}}{\|\bom(\cdot\,,\,t)\|_{1,\,loc}} < \infty\,.
\ee
\end{theorem}
Clearly, if $U_{1} = \|\bu\|_{\infty} <\infty$ and $\|\bom(\cdot\,,\,t)\|_{1,\,loc} < \infty$ 
on $[0,\,T]$ and $\rho_{0}$ is bounded away from zero then the BKM theorem says that 
no singularities can arise. The Lipschitz condition (\ref{C94c}) can be re-expressed to account 
for anti-parallel vortex tubes \cite{Const94}.
\par\smallskip
Constantin, Fefferman \& Majda \cite{CFM} then considered in more detail how to define the concept 
of ``smoothly directed'' for trajectories. Consider the three-dimensional Euler equations with 
smooth localized initial data; assume the solution is smooth on $0 \leq t < T$. The velocity field 
defines particle trajectories $\bX(\bx_{0},\,t)$ that satisfy
\bel{CFM1}
\frac{D\bX}{Dt} = \bu (\bX,\,t)\hspace{2cm}
\bX(\bx_{0},\,0) = \bx_{0}\,.
\ee
The image $\bW_{t}$ of a set $\bW_{0}$ is given by $\bW_{t} = \bX(t,\,\bW_{0})$. Then the set 
$W_{0}$ is said to be \textit{smoothly directed} if there exists a length $\rho >0$ and a 
ball $0 < r < \shalf\rho$ such that the following 3 conditions are satisfied:
\ben\itemsep -1mm
\item For every $\bx_{0} \in \bW_{0}^{*}$ where $\bW_{0}^{*} = \left\{\bx_{0} \in \bW_{0}^{*}\,;~|\bom_{0}(\bx_{0})|
\neq 0\right\}$ and all $t \in [0,\,T)$, the function $\bhom(\cdot\,,\,t)$ has a Lipschitz extension 
to the ball of radius $4\rho$ centred at $\bX(\bx_{0},\,t)$ and 
\bel{CFM3}
M = \lim_{t\to T}\sup_{\bx_{0}\,\in \bW_{0}^{*}}\int_{0}^{t}
\|\nabla\bhom(\cdot\,,\,t)\|_{L^{\infty}(B_{4\rho})}^{2}\,dt < \infty\,.
\ee
This assumption ensures the direction of vorticity is well-behaved in the neighbourhood 
of a set of trajectories.
\item The condition
\bel{CFM4}
\sup_{B_{3r}(\bW_{t})}|\bom(\bx,\,t)| \leq m\sup_{B_{r}(\bW_{t})}|\bom(\bx,\,t)|
\ee
holds for all $t \in [0,\,T)$ with $m = \hbox{const}>0$. This simply means that this chosen 
neighbourhood captures large \& growing vorticity but not so much that it overlaps with 
another similar region.  

\item The velocity field in the ball of radius $4\rho$ satisfies
\bel{CFM5}
\sup_{B_{4r}(\bW_{t})}|\bu(\bx,\,t)| \leq U(t) := \sup_{\bx}|\bu(\bx,\,t)|<\infty\,,
\ee
for all $t \in [0,\,T)$. 
\een
\begin{theorem}\label{CFMthm3} (Constantin, Fefferman \& Majda \cite{CFM}) Assume that $\bW_{0}$ 
is smoothly directed as in (i)--(iii) above. Then there exists a time $\tau>0$ and a constant 
$\Gamma$ such that 
\bel{CFM6}
\sup_{B_{r}(\bW_{t})}|\bom(\bx,\,t)| \leq \Gamma 
\sup_{B_{\rho}(\bW_{t})}|\bom(\bx,\,t_{0})|
\ee
holds for any $0 \leq t_{0} < T$ and $0 \leq t - t_{0} \leq \tau$.
\end{theorem}
Condition (ii) may have implications for how the natural length $\rho$ scales with 
time as the flow develops \cite{Kerr05} but more work needs to be done to understand its 
implications. Cordoba \& Fefferman \cite{CF} have weakened condition (iii) in the case of 
vortex tubes to
\bel{CFcrit}
\int_{0}^{T} U(s)\,ds = \int_{0}^{T}\|\bu(\cdot\,,\,s)\|_{\infty}\,ds < \infty\,.
\ee

\subsection{The work of Deng, Hou \& Yu}\label{DHYsub}

Deng, Hou \& Yu \cite{DHY1} have re-worked probably the most important of 
the ``smoothly directed criteria'', namely (\ref{CFM3}), from local control over 
$\int_{0}^{t}\|\nabla\bhom(\cdot\,,\,t)\|_{L^{\infty}}^{2}dt$ in $0 \leq t \leq T$ to a 
condition on the arc-length $s$ between two points $s_{1}$ and $s_{2}$. The first of 
their two results is\,:
\begin{theorem}\label{DHYthm1} (Deng Hou \& Yu \cite{DHY1}){\bf\,:}
Let $\bx(t)$ be a family of points such that $|\bom(\bx(t),\,t)| \gtrsim \Omega(t) 
\equiv \|\bom\|_{\infty}$. Assume that for all $t \in [0,\,T]$ there is another 
point $\by(t)$ on the same vortex line as $\bx(t)$ such that the unit vector of 
vorticity $\bhom(\bx,\,t)$ along the line between $\bx(t)$ and $\by(t)$ is well-defined. 
If we further assume that 
\bel{DHY1a}
\left|\int_{s_{1}}^{s_{2}} \hbox{div}\,\bhom(s,\,t)\,ds\right| \leq C(T)
\ee
together with
\bel{DHY2a}
\int_{0}^{T}|\bom(\bx(t),\,t)|\,dt < \infty\,,
\ee
then there will be no blow-up up to time $T$. Moreover, 
\bel{DHY2b}
e^{-C} \leq \frac{|\bom(\bx(t),\,t)|}{|\bom(\by(t),\,t)|} \leq e^{C}\,.
\ee
\end{theorem}
Inequality (\ref{DHY1a}) is based on the simple fact that 
\bel{DHY3a}
0 = \hbox{div}\,\bom = |\bom|\,\hbox{div}\,\bhom + \bhom\cdot\nabla|\bom|
= |\bom|\,\hbox{div}\,\bhom + \frac{~\,d|\bom|}{ds}
\ee
where $\bhom\cdot\nabla = \frac{d~}{ds}$ is the arc-length derivative. 
\par\smallskip
The second and more important of the results of Deng, Hou \& Yu \cite{DHY2} is based on  
considering a family of vortex line segments $L_{t}$ along which the maximum vorticity 
is comparable to the maximum vorticity $\Omega (t)$.  Denote by $L(t)$ the arc length 
of $L_{t}$, $\bhn$ the unit normal vector, and $\kappa$ the curvature of the vortex 
line. Furthermore, they define
\bel{DHY2ndthm}
U_{\hat{\omega}}(t)\equiv\max_{\scriptsize\bx,\scriptsize\by\in L_{t}}\left|\left(\bu\cdot\bhom\right)
(\bx,t)-\left(\bu\cdot\bhom\right)(\by,t)\right|\,,
\ee
\bel{DHYdef1}
U_{n}(t)\equiv\max_{L_{t}}\left|\bu\cdot \bhn\right|\,,
\ee
and
\bel{DHYdef2}
M(t)\equiv\max\left(\left
\Vert \nabla\cdot\bhom\right\Vert _{L^{\infty}(L_{t})},\left\Vert \kappa
\right\Vert _{L^{\infty}(L_{t})}\right)\,.
\ee
\begin{theorem}\label{DHYthm2}(Deng Hou \& Yu \cite{DHY1}){\bf\,:} Let $A,B\in\left(0,1\right)$ 
with $B<1-A$, and $C_0 $ be a positive constant. If
\ben\itemsep -1mm
\item $U_{\hat{\omega}}(t)+U_{n}(t)\lesssim(T-t)^{-A}$,

\item $M(t)L(t)\le C_{0}$,

\item $L(t)\gtrsim\left(T-t\right)^{B}$,
\een
then there will be no blow-up up to time $T$. 
\end{theorem}
In a further related paper Deng, Hou \& Yu \cite{DHY2} have changed the inequality $A+B <1$ to 
equality $A+B=1$ subject to a further weak condition. They also derived some improved geometric 
scaling conditions which can be applied to the scenario when the velocity blows up at the same 
time as vorticity and the rate of blow-up of velocity is proportional to the square root of 
vorticity. This is the worst possible blow-up scenario for velocity field due to Kelvin's 
circulation theorem.

\subsection{The non-constancy of $\alpha_{p}$ \& $\bchi_{p}$: quaternions \& the direction of 
vorticity}\label{direc2}

The key relation in the quaternionic formulation of the Euler equations is the Riccati 
equation (\ref{Rictriad}) for $\bfq = [\alpha(x,t),\,\bchi(x,\,t)]$. In terms of $\alpha$ 
and $\chi$ this gives four equations
\bel{ac2}
\frac{D\alpha}{Dt} = \chi^2 - \alpha^2 -\alpha_{p}\,,
\hspace{2cm}
\frac{D\bchi}{Dt} = -2\alpha\bchi - \bchi_{p}\,.
\ee
Although apparently a simple set of differential equations driven by $\bfq_{p} = [\alpha_{p},
\,\bchi_{p}]$, it is clear that $\bfq_{p}$ is not independent of the solution because of the 
pressure constraint $-Tr\,P = u_{i,k}u_{k,i}$. In consequence it is tempting to think of 
$\bfq_{p}$ as behaving in a constant fashion. This may be true for large regions of an Euler 
flow but it is certainly not true in the most intense vortical regions where vortex lines 
have their greatest curvature; in these regions the signs of $\alpha_{p}$ and 
of the components of $\bchi_{p}$ may change dramatically \cite{GGK}. It is because of these 
potentially violent changes that $\bfq_{p}$ could be considered as a candidate for a 
further conditional direction of vorticity theorem along the lines of those in \S\ref{CFMsub} 
and \S\ref{DHYsub}. Other work where constraints on $P$ appear is the paper by Chae \cite{DC6}.
\par\medskip
The work in \cite{CFM,DHY1,DHY2} shows that $\nabla\bw$ needs to be controlled in some fashion 
in local areas where vortex lines have high curvature. In terms of the number of derivatives 
the Hessian $P$ is on the same level and it is in terms of $P$ and the variables associated 
with it ($\alpha_{p}$ and $\bchi_{p}$) that we look for control of Euler solutions. 
From their definitions, it is easily shown that $\alpha^{2} + \chi^{2} = |S\bhom|^{2}$ 
and thus on vortex lines, $\alpha = \alpha(\bX(t,\,{\small\bx_{0}}),\,t)$, (\ref{ac2}) becomes
\bel{S0}
\frac{d~}{dt}|S\bhom|^{2} = - \alpha |S\bhom|^{2} + \alpha\alpha_{p} + \bchi\cdot\bchi_{p}
\ee
Thus on integration
\bel{S1}
|S\bhom(\bX(\tau),\,t)|^{2} = -2 \int_{0}^{T}e^{\int_{0}^{\tau}
\alpha(\cdot,\,t')\,dt' - \int_{0}^{t}\alpha(\cdot,\,t')\,dt'}
\left(\alpha\alpha_{p}+ \bchi\cdot\bchi_{p}\right)(\bX(\cdot,\,\tau)\,d\tau\,.
\ee
\par\medskip
There are now two alternatives. The first is to make one application of a Cauchy-Schwarz 
inequality and use the fact that $\alpha_{p}^{2} + \chi_{p}^{2} = |P\bhom|^{2}$
\bel{S2}
|S\bhom(\bX(t,\,{\small\bx_{0}}),\,t)| \leq 2 \int_{0}^{T}e^{\int_{0}^{\tau}
\alpha(\cdot,\,t')\,dt' - \int_{0}^{t}\alpha(\cdot,\,t')\,dt'}
\, |P\bhom(\cdot,\,\tau)|\,d\tau\,.
\ee
This is similar to Chae's result (his Theorem 5.1 in \cite{DC6}) which is based on control 
of the time integral of $\|S\bhom\cdot P\bhom\|_{\infty}$, which is derivable from (\ref{ertel1}).
\par\medskip
The second raises an interesting case respecting the direction of vorticity using $\bchi_{p}$ 
and can be viewed as an alternative way of looking at the direction of vorticity after 
\cite{CFM,DHY1,DHY2}. $\bchi_{p} = \bhom\times P\bhom$ contains $\bhom$ not $\bom$ \& is thus 
concerned with the direction of $\bom$ rather than its magnitude. Firstly we use the fact that 
$|\bom|$ cannot blow-up for $\alpha < 0$ because $D|\bom|/Dt = \alpha|\bom|$\,; thus our concern 
is with $\alpha \geq 0$. In the case when the angle between $\bhom$ and $P\bhom$ is not zero 
\bel{S3}
|S\bhom(\bX(t,\,{\small\bx_{0}}),\,t)| \leq 2 \int_{0}^{T}|\bchi_{p}(\cdot,\,\tau)|\,d\tau\,.
\ee
If the right hand side is bounded then Euler cannot blow up, excepting the possibility that 
$|P\bhom|$ blows up simultaneously as the angle between $\bhom$ and $P\bhom$ approaches zero 
while keeping $\bchi_{p}$ finite; under these circumstances $\int_{0}^{t}|\bchi_{p}|d\tau < \infty$,
whereas $\int_{0}^{t}|\alpha_{p}|d\tau \to\infty$ and thus blow-up is still theoretically 
possible in that case. The result does not imply that blow-up occurs when collinearity does; it
simply implies that under condition (\ref{S3}) it is the only situation when it can happen.
Ohkitani \cite{Ohk93} and Ohkitani and Kishiba \cite{Ohk95} have noted the collinearity mentioned 
above; they observed in Euler computations that at maximum points of enstrophy, $\bom$ tends to 
align with the eigenvector corresponding to the most negative eigenvalue of $P$. Expressed over 
the whole periodic volume we have\,: 
\begin{theorem}\label{chipthm}(Gibbon, Holm, Kerr \& Roulstone \cite{GHKR}){\bf\,:} 
On the domain $\mathbb{D} = [0,L]^{3}_{per}$ there exists a global solution of the Euler 
equations, $\bu \in C([0,\,\infty];H^{s})\cap C^{1}([0,\,\infty];\,H^{s-1})$ for $s\geq 3$ if, 
for every $T>0$
\bel{integ1b}
\int_{0}^{T}\|\bchi_{p}(\cdot,\,\tau)\|_{\infty}\,d\tau < \infty\,.
\ee
excepting the case where $\bhom$ becomes collinear with an eigenvector of $P$ at $T$.
\end{theorem}

\section{\large A final example: the equations of incompressible ideal MHD}\label{examples}

The Lagrangian formulation of \S\ref{lagev} can be applied to many situations, such as the 
stretching  of fluid line-elements, incompressible motion of Euler fluids and ideal MHD 
(Majda \& Bertozzi \cite{MB01}). We choose ideal MHD in Elsasser variable form as a final 
example; another approach to this can be found in \cite{GHnjp}. The equations for the fluid 
and the magnetic field $\bB$ are
\bel{F1}
\frac{D\bu}{Dt} = \bB\cdot\nabla\bB - \nabla p\,,
\ee
\bel{F2}
\frac{D\bB}{Dt} = \bB\cdot\nabla\bu\,,
\ee
together with $\mbox{div}\,\bu = 0$ and $\mbox{div}\,\bB = 0$. The pressure
$p$ in (\ref{F1}) is $p = p_{f} + \frac{1}{2}B^{2}$ where $p_{f}$ is the 
fluid pressure. Elsasser variables are defined by the combination \cite{HKM1}
\bel{F3}
\bv^{\pm} = \bu \pm \bB\,.
\ee
The existence of two velocities $\bv^{\pm}$ means that there are two material 
derivatives
\bel{F3a}
\frac{D^{\pm}}{Dt} = \frac{\partial~}{\partial t} + \bv^{\pm}\cdot\nabla\,.
\ee
In terms of these, (\ref{F1}) and (\ref{F2}) can be rewritten as 
\bel{F4}
\frac{D^{\pm}\bv^{\mp}}{Dt} = - \nabla p\,,
\ee
with the magnetic field $\bB$ satisfying ($\mbox{div}\,\bv^{\pm} = 0$)
\bel{F5}
\frac{D^{\pm}\bB}{Dt} = \bB\cdot\nabla \bv^{\pm}\,.
\ee
Thus we have a pair of triads 
$(\bv^{\pm},\,\bB,\,\ba^{\pm})$ with $\ba^{\pm} = \bB\cdot\nabla \bv^{\pm}$, 
based on Moffatt's  identification of the $\bB$-field as the important 
stretching element \cite{HKM1}. From \cite{Gibbon02,GHKR} we also have 
\bel{F8}
\frac{D^{\pm}\ba^{\mp}}{Dt} = - P\bB\,,
\ee
where $\bdb^{\pm} = -P\bB$. With two quartets $(\bv^{\pm},\,\bB,\,\ba^{\pm}\,,\bdb)$, 
the results of \S\ref{lagev} follow, with two Lagrangian derivatives and two Riccati 
equations 
\bel{Ric2}
\frac{D^{\mp}\bfq_{a}^{\pm}}{Dt} + \bfq_{a}^{\pm}\cast\bfq_{a}^{\mp} = \bfq_{b}\,.
\ee
In consequence, MHD-quaternion-frame dynamics needs 
to be interpreted in terms of two sets of ortho-normal frames $\left(\bhB,\,\bhchi^{\pm},\,
\bhB\times\bhchi^{\pm}\right)$ acted on by their opposite Lagrangian time derivatives. 
\beq\label{frameMHD1}
\frac{D^{\mp}\bhB}{Dt}&=& \bD^{\mp}\times\bhB\,,
\\
\frac{D^{\mp}}{Dt}(\bhB\times\bhchi^{\pm}) &=& \bD^{\mp}\times(\bhB\times\bhchi^{\pm})\,,
\\
\frac{D^{\mp}\bhchi^{\pm}}{Dt} &=& \bD^{\mp}\times\bhchi^{\pm}\,,
\eeq
where the pair of Elsasser Darboux vectors $\bD^{\mp}$ are defined as
\bel{frameMHD2}
\bD^{\mp} = \bchi^{\mp} - \frac{c_{B}^{\mp}}{\chi^{\mp}}\bhB\,,
\hspace{2cm}
c_{B}^{\mp} = \bhB\cdot[\bhchi^{\pm}\times(\bchi_{pB} + \alpha^{\pm}\bchi^{\mp})]\,.
\ee

\section{\large Conclusion}

The well-established use of quaternions by the aero/astro-nautics and computer animation 
communities in the spirit intended by Hamilton gives us confidence that they are 
applicable to the `flight' of Lagrangian particles in both passive tracer particle flows 
and, in particular, three-dimensional Euler flows. An equivalent formulation for the 
compressible Euler equations (\cite{jts2,jts3}) may give a clue to the nature of the 
incompressible limit \cite{EJDG}. The case of the barotropic compressible Euler equations 
and other examples are given in the summary below in Table 2\,:
$$
\begin{array}{c|lccr|c}
\hbox{System}&\quad\bu\quad & \quad\bw\quad & \quad\ba\quad & \quad\bdb\quad& 
\hbox{\small Material~Deriv}\\\hline
\hbox{\small incompressible Euler}&\quad\bu\quad & \quad\bx\quad & \quad\bu\quad & 
\quad -\nabla p \quad & D/Dt\\
\hbox{\small incompressible Euler}&\quad\bu\quad & \quad\bom\quad & \quad S\bom\quad & 
\quad-P\bom\quad& D/Dt\\
\hbox{\small barotropic Euler}&\quad\bu\quad & \quad\bom/\rho\quad & \quad (\bom/\rho)
\cdot\nabla\bu\quad & - (\omega_{j}/\rho)\partial_{j}(\rho \partial_{i}\,p)& D/Dt\\
\hbox{\small MHD}&\quad\bv^{\pm}\quad & \quad \bB\quad & \quad\bB\cdot\nabla\bv^{\mp}\quad & 
\quad-P\bB\quad& D^{\pm}/Dt\\
\hbox{\small Mixing}&\quad\bu\quad & \quad\bdl\quad & \quad \bdl\cdot\nabla\bu\quad & 
\quad- P\bdl\quad&D/Dt
\end{array}
$$
{\small Table 2\,: The entries display various examples of the use of Ertel's Theorem in closing 
the quartet of vectors $(\bu,\,\bw,\,\ba,\,\bdb)$. For ideal MHD, $D^{\pm}/Dt$ is defined in 
(\ref{F3a}).}
\par\medskip
Whenever quaternions appear in a natural 
manner, it usually a signal that the system has inherent geometric properties. For the Euler 
equations, it is significant that this entails the growth rate $\alpha$ and swing rate $\bchi$ 
of the vorticity vector, the latter being very sensitive to the direction of vorticity with 
respect to eigenvectors of $S$. To elaborate further, consider a Burgers' vortex which represents 
a vortex tube  \cite{MKO}. An eigenvector of $S$ lies in the direction of the tube-axis parallel 
to $\bom$ in which case $\bchi = \bhom\times S\bhom = 0$. However, if a tube comes into close 
proximity with another then they will bend and maybe tangle. As soon as the tube-curvature 
becomes non-zero along a certain line-length then $\bchi\neq 0$ along that length. Likewise 
this will also be true for vortex sheets that bend or roll-up when in close proximity to 
another sheet. The 3-vector $\bchi$ is therefore sensitively and locally dependent on the 
vortical topology. In fact at each point its evolution is most elegantly expressed through 
its associated quaternion $\bfq$, which must satisfy (see (\ref{Rictriad}))
\bel{Riccon}
\frac{D\bfq}{Dt} + \bfq\cast\bfq + \bfq_{p} = 0\,.
\ee
To fully appreciate the power of the method the pressure field must necessarily appear explicitly 
in the form of its Hessian through $\bfq_{p}$ although this runs counter to conventional 
practice in fluid dynamics where it is usually removed using Leray's projector. The pressure 
Hessian appears in the material derivative of the vortex stretching term, through the use of 
Ertel's Theorem, as the price to be paid for cancelling nonlinearity $O(|\bom||\nabla\bu|^2)$.
In fact, the effect of the pressure Hessian on the vorticity stretching term is subtle and 
non-local. Therefore, while it is tempting to discount the pressure because it disappears 
overtly in the equation for the vorticity, covertly it may arguably be one of the most 
important terms in inviscid fluid dynamics. 

\par\smallskip
There are, of course, stationary solutions of (\ref{Riccon}) one of which is $\bchi = \bchi_{p}=0$
with $\alpha = \alpha_{0}$ and $\alpha_{p} = -\alpha_{0}^{2}$. The Burgers' vortex is a solution 
of this type: see \cite{GGK,Gibbon02}.  Having laid much stress in \S\ref{direc2} on the non-constancy
of $\alpha_{p}$ and $\bchi_{p}$ in intense, potentially singular regions, nevertheless let us to 
try to determine the simplest generic behaviour of $\alpha$ and $\bchi$ from (\ref{ac2}) when 
$\alpha_{p}$ and $\bchi_{p}$ are constant; for example, a near-Burgers' vortex. To do this let us 
consider the four equations  which come out of (\ref{Riccon}), as in (\ref{ac2}), and think of 
them as ordinary differential equations on particle paths 
$\bX(t,{\small\bx_{0})}$ 
\bel{alchiextra1}
\dot{\alpha} = \chi^{2}-\alpha^{2}-\alpha_{p}\,,
\hspace{2cm}
\dot{\chi} = -2\alpha\bchi - C_{p}\,.
\ee
In regions of the $\alpha-\chi$ phase plane where $\alpha_{p} =const,~~C_{p} =\bhchi\cdot\bchi_{p}
= const$ there are 2 critical points
\bel{alchiextra2}
(\alpha,\,\chi) = (\pm\alpha_{0},\,\chi_{0})\hspace{2cm}
2\alpha_{0}^2 = \alpha_{p} + [\alpha_{p}^2 + C_{p}^2]^{1/2}
\ee
Thus there are two fixed points; one with $\alpha_{0} > 0$ (stretching), which is a stable spiral, 
and one with $\alpha_{0} < 0$ (compression); both have a small and equal value of $\chi_{0}$. The 
point with $\alpha_{0} < 0$ is an unstable spiral while $\alpha_{0} > 0$ is stable. 
Perhaps it is a surprise that it is the stretching case that is the
attracting point although it should also be noted that these equations without the Hessian terms 
have arisen in Navier-Stokes turbulence modelling \cite{YiMen}.
\par\smallskip
Finally, the existence of the relation (\ref{Riccon}), and its more general Lagrangian equivalent 
(\ref{Rictriad}), is the key step in proving Theorem \ref{abthm}, from which the frame dynamics is 
derived. Moreover, for the three-dimensional Euler equations, (\ref{Riccon}) is also the key step 
in the proof of Theorem \ref{chipthm}. 

\par\vspace{2mm}\noindent
\textbf{Acknowledgements:} {\small For discussions I would like to thank Darryl Holm, Greg 
Pavliotis, Trevor Stuart, Arkady Tsinober and Christos Vassilicos, all of Imperial College 
London, Uriel Frisch of the Observatoire de Nice, Tom Hou of California Institute of Technology, 
Robert Kerr of the University of Warwick, Ian Roulstone of the University of Surrey, Edriss Titi 
of the Weizmann Institute, Israel and Vladimir Vladimirov of the University of York. For their 
kind hospitality I would also like to thank the organizers 
(particularly Professors Andrei Fursikov and Sergei Kuksin) of the Steklov Institute's 
\textit{Mathematical Hydrodynamics} meeting, held in Moscow in June 2006.}


\bibliographystyle{unsrt}

\end{document}